\documentclass[12pt]{iopart}
\usepackage{iopams}
\usepackage{epsfig,graphicx}
\begin{document}

\title[Relaxation of Electron Spin during High-Field Transport in GaAs Bulk]{Relaxation of Electron Spin during High-Field Transport in GaAs Bulk}

\author{S. Spezia$^a$, D. Persano Adorno$^a$, N. Pizzolato$^{ab}$, B. Spagnolo$^{ab}$}
\address{$^a$Dipartimento di Fisica e Tecnologie Relative, \\
Universit\`a di Palermo and CNISM,\\
Viale delle Scienze, edificio 18, I-90128 Palermo, Italy\\
$^b$Group of Interdisciplinary Physics} \ead{stefano.spezia@gmail.com}

\begin{abstract}
A semiclassical Monte Carlo approach is adopted to study the
multivalley spin depolarization of drifting electrons in a doped
n-type GaAs bulk semiconductor, in a wide range of lattice
temperature ($40<T_L<300$ K) and doping density
($10^{13}<n<10^{16}$cm$^{-3}$). The decay of the initial
non-equilibrium spin polarization of the conduction electrons is
investigated as a function of the amplitude of the driving static
electric field, ranging between $0.1$ and $6$ kV/cm, by considering
the spin dynamics of electrons in both the $\Gamma$ and the upper
valleys of the semiconductor. Doping density considerably affects
spin relaxation at low temperature and weak intensity of the driving
electric field. At high values of the electric field, the strong
spin-orbit coupling of electrons in the $L$-valleys significantly
reduces the average spin polarization lifetime, but, unexpectedly,
for field amplitudes greater than 2.5 kV/cm, the spin lifetime
increases with the lattice temperature. Our numerical findings are
validated by a good agreement with the available experimental
results and with calculations recently obtained by a different
theoretical approach.
\end{abstract}

\maketitle

\section{Introduction}\label{sect1}
The processing of a high volume of information and world wide
communication is, at the present, based on semiconductor technology,
whereas information storage devices rely on multilayers of magnetic
metals and insulators. Semiconductor spintronics offers a possible
direction of technological research towards the development of
hybrid devices that could perform logic operations, communication
and storage, within the same material technology: information could
be stored in a system of polarized electron spins
\cite{FabianDasSarma1999,Wolf2001,Awschalom2007,Fabian2007,Lou07,Jon07,DyakonovEd,Gul09,Flatte09,Dash2009,WuReport2010},
transferred as attached to mobile carriers and finally detected. The
possibility of obtaining long spin relaxation times or spin
diffusion lengths in electronic materials makes spintronics a viable
prospective technology. Nevertheless, the designers of spin devices
have to worry about the loss of spin polarization (spin coherence)
before, during and after the necessary manipulations. In particular,
efficient injection, transport, control and detection of spin
polarization must be carefully treated~\cite{Wolf2001}.
Electron-spin states depolarize by scattering with imperfections or
elementary excitations of the medium, such as phonons. Furthermore,
miniaturization process brings the system to experience very intense
electric fields, even when the applied voltages are very low. This
means that, for the operability of prospective spintronic devices,
the features of spin relaxation at relatively high electric fields
should be firstly understood. In recent years there was a
proliferation of experimental works in which the influence of
transport conditions on relaxation of spins in semiconductors has
been
investigated~\cite{Hagele98,Sanada2002,Sato04,Stephens04,Crooker05,Furis2006,Beck2006}.
All these works are focused on the study of coherent spin transport
at low temperatures ($T_L<30$ K) and under the influence of weak
electric fields ($F<0.1$ kV/cm), except for few
works~\cite{Hagele98,Sanada2002} in which spin depolarization has
been investigated with driving fields up to $6$ kV/cm.

The temporal evolution of the spin and the evolution of the momentum
of an electron cannot be separated. The spin depolarization rates
are functionals of the electron distribution function in momentum
space which continuously evolves with time when an electric field is
applied to drive the transport. Thus, the dephasing rate is a
dynamic variable that needs to be treated self-consistently in step
with the dynamic evolution of the electron's momentum. A way to
solve this problem is to describe the transport of spin polarization
by making use of Boltzmann-like kinetic equations. This can be done
within the density matrix approach \cite{Ivchenko1990}, methods of
nonequilibrium Green's functions, as the microscopic kinetic spin
Bloch equation approach
\cite{Wu2000,Weng2003,Weng2004,Zhang08,Jiang09,WuReport2010}, or
Wigner functions \cite{Mishchenko2003,Saikin2004}, where spin
property is accounted for starting from quantum mechanics equations.

Another way is to use a semiclassical Monte Carlo approach, by
taking into account the spin polarization dynamics with the
inclusion in the code of the precession mechanism of the spin
polarization vector~\cite{Kiselev2000,Bournel2000, Barry2003,
Saikin2003,Pramanik2003,Shen2004,Pershin2005,Saikin2006}.

Both methods allow to include the relevant spin relaxation phenomena
for electron systems and take into account the details of electron
scattering mechanisms, material properties and specific device
design; their predictions have been demonstrated to be in good
agreement with experiments.

Theoretical descriptions of the transport of spin-polarized
electrons have been also achieved by the drift-diffusion
approximation. The existing drift-diffusion schemes can be
classified into two approaches accounting the spin degree of freedom
differently: the two-component drift-diffusion model and the
density-matrix based approximations
\cite{Yu2002,Martin2003,Hruska2006}. General conditions for the
applicability of these approximations are not different from the
usual conditions of applicability of drift-diffusion approximations.

Despite decades of studies, most of theoretical or simulative works
have considered only the central valley $\Gamma$ since the
spin-orbit coupling parameters of the upper conduction bands have
been only recently theoretically calculated by Fu et
al.~\cite{Fu2008}. Monte Carlo approaches have been widely adopted
by groups of scientists to study spin polarized transport in 2D
channels, heterostructures, quantum wells, quantum
wires~\cite{Kiselev2000,Bournel2000,Saikin2003,Pramanik2003,Shen2004,Pershin2005}.
However, till today, to the best of our knowledge, in semiconductor
bulk structures a theoretical investigation of the influence of
transport conditions on the spin depolarization in the presence of
high electric fields, comprehensive of the effects of both
lattice temperature and impurity density,  is still lacking.\\
\indent Aim of this work is to numerically estimate the spin
lifetimes of an ensemble of initially-polarized electrons drifting
in doped n-type GaAs bulks, with the lattice temperature, doping
density and electric field amplitude ranging in a wide interval of
values, focusing on the effects due to the inclusion of
upper valleys of the semiconductor on the mechanism of depolarization.\\
The paper is organized as follows: in Sec.~\ref{sect2} we briefly
describe the multivalley model, used for the study of the spin
depolarization dynamics and the Monte Carlo simulator; in
Sec.~\ref{sect3} the numerical results are given and discussed.
Final comments and conclusions are given in Sec.~\ref{sect4}.

\section{Theory and Monte Carlo approach}\label{sect2}

\subsection{Spin dynamics and multivalley model}\label{subsect21}
\indent Spin dephasing may be caused by interactions with local
magnetic fields originating from nuclei and spin-orbit interactions
or magnetic impurities. The most relevant spin relaxation mechanisms
for an electron system under non degenerate regime are: (i) the
Elliott-Yafet (EY) mechanism, in which electron spins have a small
chance to flip during each scattering, due to the spin mixing in the
conduction band \cite{Elliott54,Yafet63}; (ii) the Dyakonov-Perel
(DP) mechanism, based on the spin-orbit splitting of the conduction
band in non-centrosymmetric semiconductors, in which the electron
spins decay due to their precession around the \textbf{k}-dependent
spin-orbit fields (inhomogeneous broadening) during the free flight
between two successive scattering events
\cite{Perel1971,Dyakonov2006,DyakonovEd}; (iii) the Bir-Aronov-Pikus
(BAP) mechanism, in which electrons exchange their spins with holes
\cite{BAP1976}. Hyperfine interaction is another mechanism, usually
important for spin relaxation of localized electrons, but
ineffective in metallic regime where most of the carriers are in
extended states \cite{Abragam61,Paget77,Pikus89}.

\indent Previous theoretical \cite{Jiang09,WuReport2010} and
experimental \cite{Litvi2010} investigations indicate that the the
EY mechanism is totally irrelevant on electron spin relaxation in
n-type III-V semiconductors. Hence, in this work we analyze the spin
depolarization of drifting electrons in n-type GaAs semiconductors
by considering only the D'yakonov-Perel process.

By following the semiclassical formalism, the term of the
single electron Hamiltonian which accounts for the spin-orbit
interaction can be written as
\begin{equation}
H_{SO} = \frac{\hbar}{2}\vec{\sigma}\cdot\vec{\Omega}.
\label{HamiltonianSO}
\end{equation}
It represents the energy of electron spins precessing around an
effective magnetic field [$\vec{B}=\hbar\vec{\Omega}/\mu_Bg$] with
angular frequency $\vec{\Omega}$, which depends on the orientation
of the electron momentum vector with respect to the crystal axes.
Near the bottom of each valley, the precession vector can be written
as ~\cite{Pikus89,Fu2008}
\begin{equation}
\vec{\Omega}_{\Gamma}=\beta_{\Gamma}[k_{x}(k_{y}^{2}-k_{z}^{2})\hat{x}+k_{y}
(k_{z}^{2}-k_{x}^{2})\hat{y}+k_{z}(k_{x}^{2}-k_{y}^{2})\hat{z}]
\label{effectivefieldgammavalley}
\end{equation}
in the $\Gamma$-valley, and
\vskip-0.7cm
\begin{equation}
\vec{\Omega}_{L}=\frac{\beta_{L}}{\sqrt{3}}[\hat{x}(k_{y}-k_{z})+\hat{y}(k_{z}-k_{x})+\hat{z}(k_{x}-k_{y})]
\label{effectivefieldLvalley}
\end{equation}
in the L-valleys, located along the [111] direction of the
crystallographic axes.  In equations
(\ref{effectivefieldgammavalley})-(\ref{effectivefieldLvalley}),
$k_{i}$ ($i=x,y,z$) are the components of the electron wave vector.
$\beta_{\Gamma}$ and $\beta_{L}$ are the spin-orbit coupling
coefficients, crucial parameters for the simulation of spin
polarization. Here, we assume $\beta_{L}$=$0.26$
eV$/${\AA}$\cdot2/\hbar$, as recently theoretically estimated
~\cite{Fu2008}. In $\Gamma$-valley we consider the effects of
nonparabolicity on the spin-orbit splitting by using \cite{Pikus89},
\begin{equation}
\beta_{\Gamma}=\frac{\alpha\hbar^{2}}{m\sqrt{2m
E_{g}}}\left(1-\frac{E(\vec{k})}{E_{g}}\frac{9-7\eta+2\eta^{2}}{3-\eta}\right)
\label{betaGammavalley}
\end{equation}
where $\alpha=0.029$ is a dimensionless material-specific parameter,
$\eta = \Delta/(E_{g}+\Delta)$, with $\Delta=0.341$ eV the
spin-orbit splitting of the valence band, $E_{g}$ is the energy
separation between the conduction band and valence band at the
$\Gamma$ point, $m$ the effective mass and $E(\vec{k})$ the electron
energy.

The quantum-mechanical description of electron spin evolution is
equivalent to that of the classical momentum $\vec{S}$ experiencing
the effective magnetic field, as described by the equation of motion
\begin{equation}
\frac{d\vec{S}}{dt}=\vec{\Omega}\times\vec{S}. \label{Poisson}
\end{equation}
Every scattering event changes the orientation of the effective
magnetic field $\vec{B}$ (that strongly depends on $\vec{k}$) and
the direction of the spin precession axis.

\subsection{The Monte Carlo algorithm}\label{subsect22}
\indent The Monte Carlo approach is one of the most powerful methods
to simulate the transport properties in  semiconductor devices
beyond the quasi-equilibrium approximations
\cite{Jacoboni,Tomizawa,Moglestue}. In fact, owing to its
flexibility, the Monte Carlo method presents the remarkable
advantage of giving a detailed description of the particle motion in
the semiconductor by taking into account the details of collisions
with impurities, phonons, etc., specific device design and material
parameters, and allows us to obtain all the needed information, such
as the average velocity of electrons, temperature, current density,
etc., directly without the need of calculating first the electron
distribution function. In simulations, between two successive
scattering events, each electron propagates along a classical
trajectory and, according to the classical equations of motion, it
is affected by the presence of external fields. The time interval
between two collisions (time of free flight), the scattering
mechanisms, the collisional angle, and all the parameters of the
problem are chosen in a stochastic way, making a mapping between the
probability density of the given microscopic process and a uniform
distribution of random numbers.

In our code the conduction bands of GaAs are represented by the
$\Gamma$-valley and by four equivalent L valleys.  We do not
consider the $X$-valleys because, even for the highest value of the
driving field ($F$= 6 kV/cm), the percentage of electrons in these
valleys is always lower than $0.1\%$.  The algorithm includes: (i)
the intravalley scattering with acoustic phonons, ionized
impurities, acoustic piezoelectric phonons, polar optical phonons,
and for the the L-valleys also the scattering with optical nonpolar
phonons; (ii) the intervalley scattering with the optical nonpolar
phonons among the two valleys. The complete set of n-type GaAs
parameters used in our calculations is listed in Ref.
\cite{Persano2000}. The scattering probabilities are calculated by
the Fermi Golden Rule and assumed to be both field and spin
independent; accordingly, the influence of the external fields is
only indirect through the field-modified electron velocities.
Nonlinear interactions of the field with the lattice and bound
carriers are neglected. We also neglect electron-electron
interactions and consider electrons to be independent
\cite{Kiselev2000}. The spin polarization vector is included into
the Monte Carlo algorithm and calculated for each free carrier. From
Eq.~(\ref{Poisson}), the Monte Carlo (MC) simulator calculates the
electron spin precession, by taking into account the
scattering-induced deviations of precession vector suffered after
each collision.

All simulations are performed by using a temporal step of $10$ fs
and an ensemble of $5\cdot10^{4}$ electrons to collect spin
statistics. We assume that all donors are ionized and that the free
electron concentration is equal to the doping concentration $n$.

\section{Numerical Results}\label{sect3}

\subsection{Spin lifetime calculation}\label{subsect31}
The initial non-equilibrium spin polarization decays with time as
the electrons, driven by a static electric field,  move through the
medium, experiencing elastic and anelastic collisions. Since
scattering events randomize the direction of $\vec{\Omega}$, during
the motion, the polarization vector of the electron spin experiences
a slow angular diffusion. The dephasing of each individual electron
spin produces a distribution of spin states that results in an
effective depolarization, which is calculated by ensemble-averaging
over the spin of all the electrons.

The simulation of the spin relaxation starts with all the electrons
of the ensemble initially polarized $(\langle \vec{S}\rangle=1)$
along the $\hat{x}$-axis at the injection plane $(x_{0}=0)$. After a
transient time of typically $10^{4}$ time steps, long enough to
achieve the steady-state transport regime, the electron spins are
initialized, the spin relaxation begins and the quantity $\langle
\vec{S}\rangle$ is calculated as a function of time. In order to
extract the characteristic time $\tau$ of the spin relaxation, the
obtained trend of the spin dephasing is fitted by the following
exponentially time decaying law
\begin{equation}
\langle S_{x}\rangle (t) = A\cdot exp(-t/\tau), \label{exponential}
\end{equation}
with $A$ a normalization factor.
\begin{figure}[htbp]
\includegraphics[height=7cm,width=11cm]{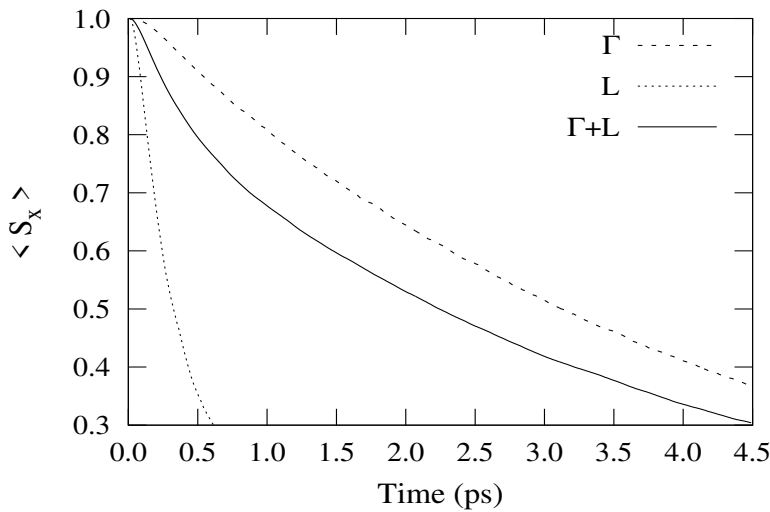}
\caption{Average electron spin polarization $\langle S_{x}\rangle$
as a function of time, by only considering the electrons drifting
into the $\Gamma$-valley (dashed line), into the $L$-valleys (dotted
line) and into the $(\Gamma+L)$-valleys (solid line). $F=5$ kV/cm,
$n=10^{13}$ cm$^{-3}$  and $T_{L}=300$ K.} \label{Fig:1}
\end{figure}

\begin{figure}[htbp]
\includegraphics[height=7cm,width=12cm]{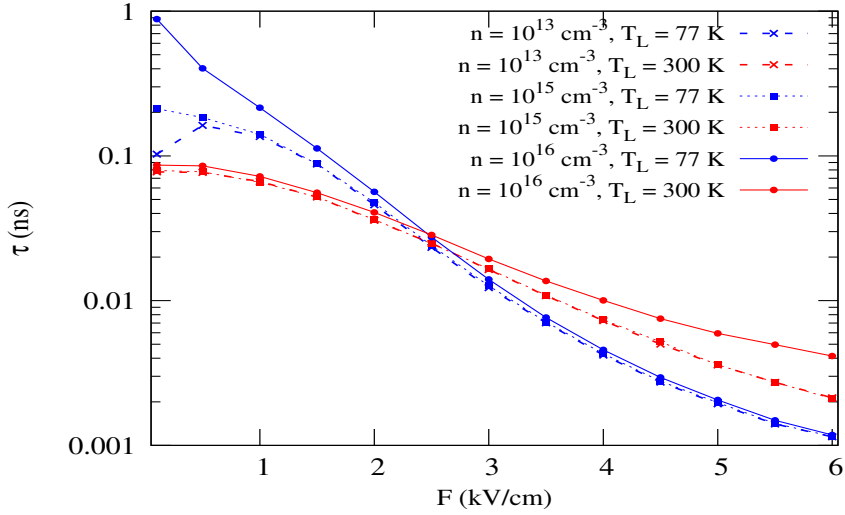}
\caption{Spin lifetime $\tau$ as a function of the electric field
amplitude $F$, at $T_L$=$77$ K (blue lines) and  $T_L$=$300$ K (red
lines), for three values of doping density, namely $n=10^{13}$
cm$^{-3}$, $n=10^{15}$ cm$^{-3}$ and $n=10^{16}$ cm$^{-3}$, with the
electrons drifting in both the $\Gamma$ and the $L$-valleys.}
\label{Fig:2}
\end{figure}

\begin{figure*}[htbp]
\includegraphics[height=5cm,width=8cm]{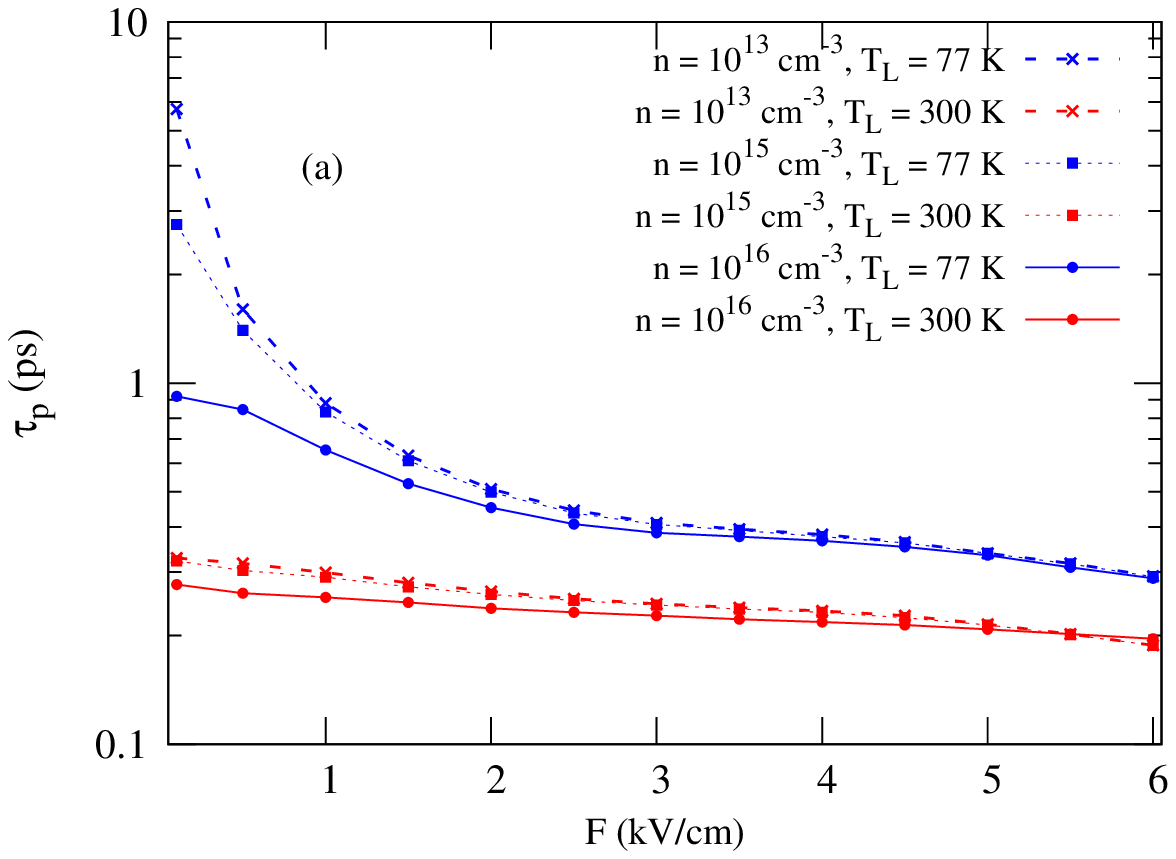}
\includegraphics[height=5cm,width=8cm]{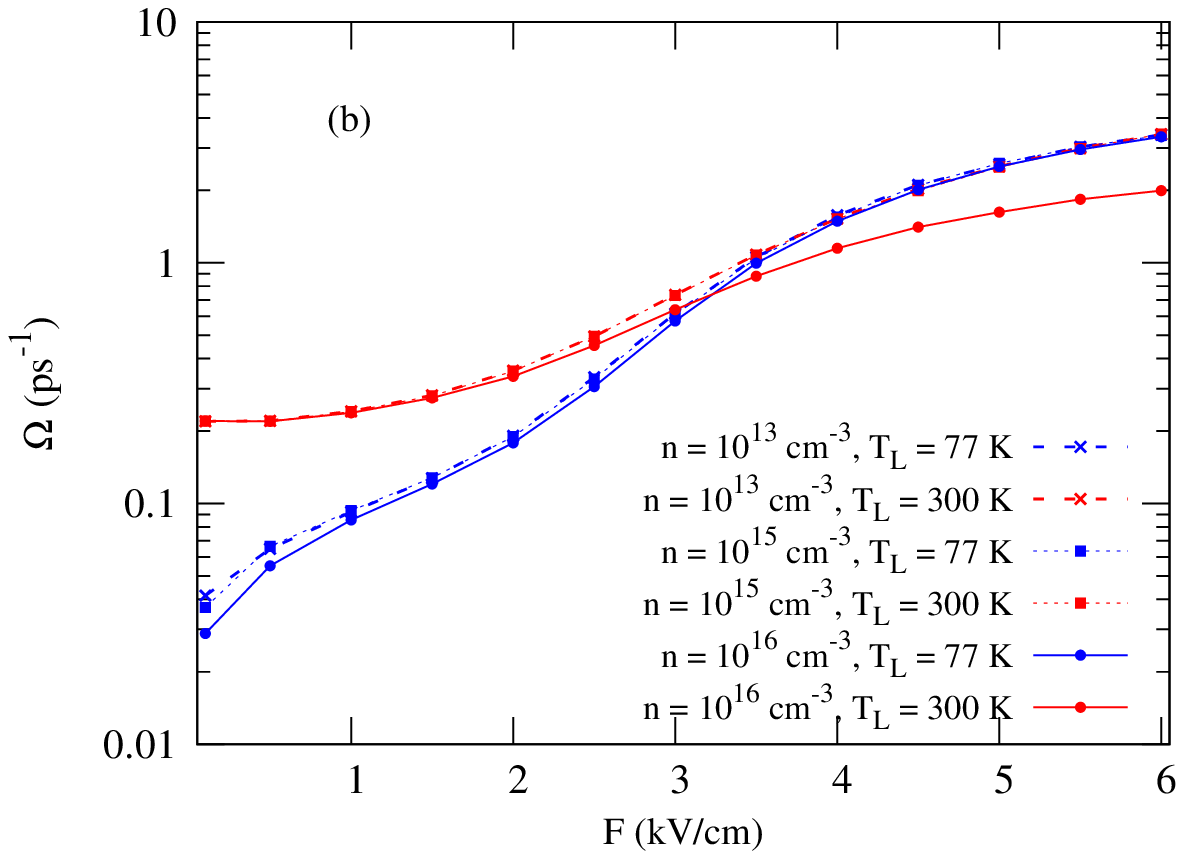}
\caption{(a) Momentum scattering time $\tau_p$ and (b) spin
precession frequency $\Omega$ as a function of the electric field
amplitude $F$, at $T_L=77$ K (blue lines) and  $T_L=300$ K (red
lines), for three values of doping density, namely $n=10^{13},
10^{15}$ and $10^{16}$ cm$^{-3}$.} \label{Fig:3}
\end{figure*}

In Fig.~\ref{Fig:1} we show the average electron spin polarization
$\langle S_{x}\rangle$, in the presence of a driving electric field,
with amplitude $F=5$ kV/cm and directed along the $\hat{x}$-axis,
with density $n= 10^{13}$ cm$^{-3}$ and lattice temperature
$T_{L}=300$ K. This value of field amplitude is high enough to allow
almost $21\%$ of all electrons to visit the $L$ valleys. The curves
represent the decreasing trend of $\langle S_{x}\rangle$ vs.~time by
firstly considering only the electrons drifting into the
$\Gamma$-valley (dashed line), secondly, by solely taking into
account the electrons moving into the $L$-valleys (dotted line) and,
finally, by considering the electrons moving into both the $\Gamma$
and the $L$-valleys (solid line). We find a significant reduction of
the average spin polarization lifetime caused by the spin-orbit
coupling in $L$-valleys stronger with respect to that in
$\Gamma$-valley, according to the theoretical results obtained by
Zhang et al.~\cite{Zhang08} on quantum wells. The transition of
about $21\%$ of electrons to the $L$-valleys leads to an increase of
efficacy of the dephasing mechanism, which brings to a reduction of
$\langle S_{x}\rangle$ over time in the range $15\div20 \%$

\begin{figure*}[ht]
\includegraphics[height=8cm,width=14cm]{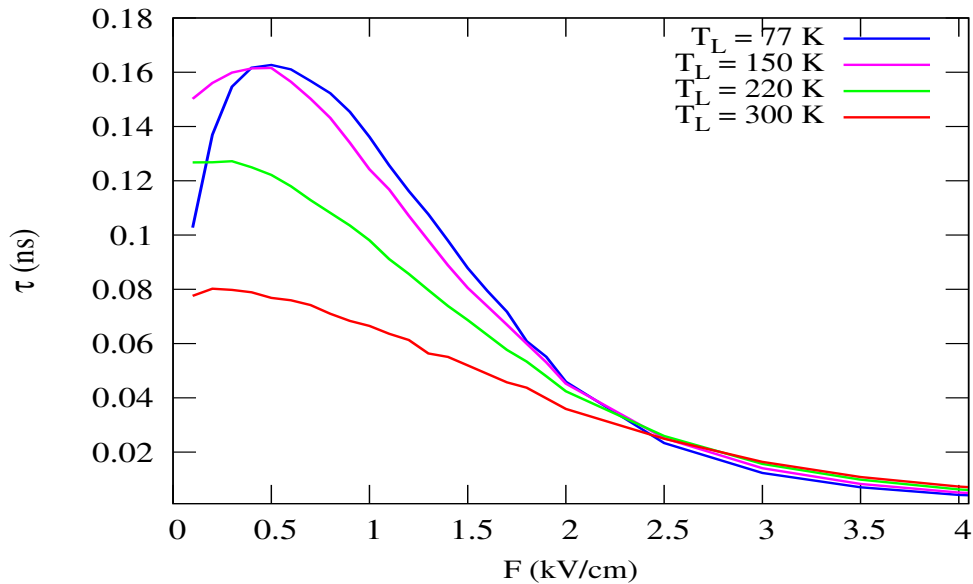}
\caption{Spin lifetime $\tau$ as a function of the electric field
amplitude $F$, at $n=10^{13}$ cm$^{-3}$ for different values of
lattice temperature, namely $T_L$=$77, 150, 220, 300$ K, with
electrons drifting in both the $\Gamma$ and the $L$-valleys.}
\label{Fig:4}
\end{figure*}

\subsection{Effects of temperature and doping density on spin relaxation}\label{subsect32}
\label{Calculation3}

In Fig.~\ref{Fig:2}, we show the spin depolarization time $\tau$ as
a function of the electric field amplitude $F$, for two values of
the lattice temperature, namely $T_L=77$ (blue curves) and $300$ K
(red curves) and three values of doping density $n=10^{13}, 10^{15}$
and $10^{16}$ cm$^{-3}$, respectively, leaving the electrons free to
drift in both the $\Gamma$ and the $L$-valleys. Except for the case
at the doping density $n=10^{13}$ cm$^{-3}$ and $T_L=77$ K, we find
that $\tau$ is always a monotonic decreasing function of $F$. In
fact, when the field amplitude becomes larger, the electron momentum
$\bf{k}$ increases, causing a stronger effective magnetic field, as
expected by Equations
(\ref{effectivefieldgammavalley})-(\ref{effectivefieldLvalley}).
Consequently, the electron precession frequency becomes higher,
inducing a faster spin relaxation \cite{Perel1971}.

For field amplitudes greater than $2.5$ kV/cm, at $T_L=300$ K we
find depolarization times longer than those obtained at $T_L=77$ K.
In order to avoid that the observed behaviour could be ascribed only
to stochastic fluctuations of MC computations, we have calculated
the statistical error associated to our simulated data. We have
repeated our simulations ten times, finding a maximum spread of
$0.05$ ps, which corresponds to about $1\%$ of the observed
variation of $\tau$ with the temperature.

We have investigated the counterintuitive behavior of longer average
spin lifetimes obtained for hotter electrons, by adopting the
proportionality law of Perel~\cite{Perel1971}:
\begin{equation}
\frac{1}{\tau}\propto \Omega^2\tau_p, \label{taus}
\end{equation}
where $\Omega$ is the spin precession frequency and $\tau_p$ the
momentum characteristic relaxation time, corresponding to the time
scale of the scattering events. We have calculated the spin
precession frequency and the momentum characteristic scattering time
for each electron of our ensemble and in Fig.~\ref{Fig:3}, we show
the average momentum scattering time $\tau_p$ (panel (a)) and the
average spin precession frequency $\Omega$ (panel (b)) as a function
of the electric field amplitude $F$, at $T_L=77$ and $300$ K, for
the three values of doping density $n=10^{13}, 10^{15}$ and
$10^{16}$ cm$^{-3}$.

The panel (a) of Fig.~\ref{Fig:3} shows that $\tau_p$ is a
monotonically decreasing function of $F$ for every values of $n$ and
$T_L$.  At the higher temperature ($T_L=300$ K) we find low values
of $\tau_p$, because electrons experience a greater number of
scattering events, both in the $\Gamma$-valley and in $L$-valleys.
Moreover, the curves at room temperature are characterized by only a
slight slope, because in this case the thermal energy of the
electrons is dominant with respect to the drift kinetic energy. At
$T_L=77$ K and for very low values of the electric field amplitude,
since the scattering events are mainly due to ionized impurities,
$\tau_p$ is greatly dependent on $n$, increasing its value at lower
densities. At $T_L=300$ K, $\tau_p$ is nearly independent on the
doping density since the dominant scattering mechanism is due to the
optical phonons.

The panel (b) of Fig.~\ref{Fig:3} shows that the spin precession
frequency $\Omega$ is an increasing monotonic function of $F$. For
$F<3$ kV/cm, independently from the values of the doping density,
the values of $\Omega$ obtained at room temperature are larger than
those obtained at $T_L=77$ K. The increase of the spin precession
frequency for electrons moving at higher temperatures is explained
by the increasing number of electron transitions from the $\Gamma$
to $L$ valleys, being the value of the spin-orbit coupling
coefficient in the $L$-valleys one order of magnitude greater than
that of $\Gamma$-valley. At $T_L=77$ K, for $F<3$ kV/cm, the
percentage of electrons in the central valley $\Gamma$ is
practically unitary and the spin precession frequency increases as
the third power of electron momentum $\vec{k}$, which increases with
$F$ according to the Eq.~(\ref{effectivefieldgammavalley}). When $F$
is greater than $3$ kV/cm, the percentage of electrons in the
L-valley is high enough to lead $\Omega$ for having a nearly linear
trend (see Eq.~(\ref{effectivefieldLvalley})). At $T_L=300$ K, for
$F<3$ kV/cm, the term of thermal energy is dominant with respect to
the drift kinetic energy and $\Omega$ vs.~$F$ shows a more slight
increase.

For $F>3$ kV/cm, independently from the values of $T_L$ and $n$, the
action of $F$ wins on the disorder due to the lattice temperature.
In fact, except for the data obtained at $T_L=300$ K and $n=10^{16}$
cm$^{-3}$, which show lower values of $\Omega$, all curves coincide.
The detached curve $T_L=300$ K and $n=10^{16}$ cm$^{-3}$ is ascribed
to a strong reduction of the percentage of electrons present in the
L-valleys.

The calculation of the square of the spin precession frequency times
the momentum relaxation time as a function of the electron energy
for each electron of the ensemble shows that for $F>2.5$ kV/cm the
average value of $\Omega^2\tau_p$ obtained at $T_L=77$ K is greater
than that at $T_L=300$ K. This finding explains the longer lifetimes
observed at higher temperatures for field amplitudes greater than
$2.5$ kV/cm .

To highlight on the nonmonotonic electric field dependence of
$\tau$, observed at the doping density $n=10^{13}$ cm$^{-3}$, we
have investigated the spin lifetime dependence on $F$ also for
different values of the lattice temperature, namely $T_L$=$77, 150,
220, 300$ K (see Fig.~\ref{Fig:4}). Up to $T_L=150$ K the
decoherence times slightly depend on the temperature and are
characterized by a marked maximum. The presence of a maximum in the
spin depolarization time can be explained by the interplay between
two competing factors, both due to the increase of the electric
field. In the momentum space, at greater field amplitudes, the
electrons occupy states with larger $\vec k$, characterized by a
stronger spin-orbit coupling, causing an enhancement of the spin
inhomogeneous broadening. On the other hand, a larger electric field
also brings about an increase of the number of scattering events,
giving rise to a reduction of the momentum relaxation time. This in
turn causes an increase of the spin relaxation time as follows from
Eq.~\ref{taus}.

At low values of temperature and for electric field amplitudes
$0.1\leq F\leq0.5$ kV/cm the inhomogeneous broadening is still
marginal and the spin relaxation phenomenon is dominated by the
momentum scattering. In particular, the number of electron
scattering events, which are mainly due to interactions with
acoustic phonons at very weak electric fields, increases its value
because of the triggering of the scattering mechanism by ionized
impurities, causing a reduction of $\tau_p$. For field amplitudes
greater than $\approx0.5$ kV/cm, the enhancement of the spin-orbit
coupling, which is k-cubic dependent, is faster than the decrease on
$\tau_p$. Consequently, the spin lifetime starts to decrease with
the increasing of the electric field. The nonmonotonic electric
field dependence of $\tau$ is not observed for $T_L>150$ K where,
because of the greater drift electron velocities, the loss of spin
polarization is mainly due to the strong effective magnetic field.

\subsection{Comparison with experiments and with other theoretical approaches}\label{subsect33}
Unfortunately, till today experimental investigations of the
ultrafast relaxation of electron spin during drift transport in bulk
semiconductors, at both sample temperatures higher than 30 K and
applied field amplitudes greater than 0.1 kV/cm, are still missing.

\begin{figure}[ht]
\includegraphics[height=7cm,width=11cm]{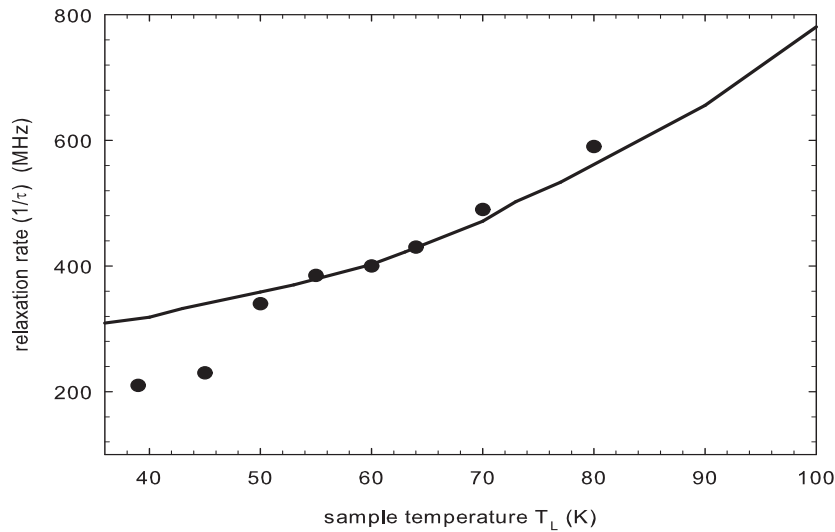}
\caption{Temperature depended measurements of the spin-relaxation
rate from the experiment (sample B) in Ref.~\cite{Romer2010}
(circles) and numerical data obtained from our Monte Carlo code
(solid curve), $n=2.7\cdot10^{15}$ cm$^{-3}$, $\beta_{\Gamma}$=$19$
eV$\cdot$\AA$^{3}$} \label{Fig:5}
\end{figure}

Although the main aim of this work is the investigation of the
influence of transport conditions on the spin relaxation of
electrons driven by high-intensity electric fields, including the
effects of choosing both the lattice temperature and impurity
density in a wide range of values, in order to validate the
prediction capability of our MC code, we have performed a comparison
between our numerical spin relaxation times and very recent
experimental results on the electron-spin-relaxation rate, obtained
by Ref.~\cite{Romer2010}. These experiments were carried out by
performing spin noise spectroscopy on a sample of n-type GaAs at a
doping concentration of $n=2.7\cdot10^{15}$ cm$^{-3}$, without any
driving field and for lattice temperatures $T_L$ between 4 and 80 K.
In Fig.~\ref{Fig:5} we plot the temperature dependence of the
spin-relaxation rate calculated from our code (solid line), together
with the experimental data (circles, sample B in
Ref.~\cite{Romer2010}). In order to best fit the experimental points
with our numerical trend, we have utilized the spin-orbit coupling
coefficient in $\Gamma$-valley $\beta_{\Gamma}$ as a free parameter,
obtaining the best agreement with $\beta_{\Gamma}$=$19$
eV$\cdot$\AA$^{3}$. This value is only slightly different from the
value ($23.8$ eV$\cdot$\AA$^{3}$), recently estimated by using the
tight binding theory~\cite{Fu2008}. However this value is still
within the reasonable range of values calculated and measured via
various methods, as reported in Ref.~\cite{Krich2007}. For sample
temperatures greater than $45$ K, our numerical trend well agrees
with the experimental data, while at lower temperatures a
considerable discrepancy is found, probably due to the neglecting
the electron-electron scattering mechanism. In fact, as shown in
Ref.~\cite{Jiang09} at $T_L$=$40$ K, the assumption of neglecting
the Coulomb term gives spin lifetime values smaller (hence
relaxation rates higher) than those obtained from full calculation
including all scattering mechanisms.

\vspace{0.5cm}

\begin{figure*}[ht]
\includegraphics[height=7cm,width=11cm]{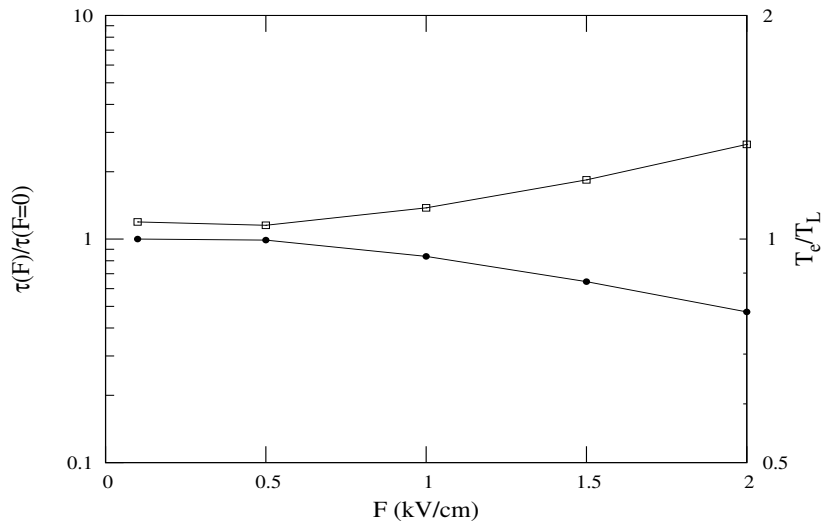}
\caption{Ratio of the spin relaxation time under electric field to
the electric-field-free one $\tau(F)/\tau(F=0)$ and ratio between
the hot-electron temperature and the lattice temperature $T_e/T_L$
as a function of the applied field obtained from our Monte Carlo
code, $n=10^{16}$ cm$^{-3}$, $T_L=300$ K.} \label{Fig:6}
\end{figure*}

In order to further test the effectiveness of our code we have
compared our one-valley numerical data with the calculation of the
effects of a low-amplitude electric field ($F\leq2$ kV/cm) on spin
relaxation in n-type III-V semiconductor bulks, lately obtained from
the fully microscopic kinetic spin Bloch equation (KSBE) approach
\cite{Jiang09}. To the best of our knowledge, that paper is the only
one in which the electric field dependence of spin lifetime has been
investigated. In Fig.~\ref{Fig:6} we plot the ratio of the spin
relaxation time under electric field to the electric-field-free one
$\tau(F)/\tau(F=0)$ and the ratio between the hot-electron
temperature and the lattice temperature $T_e/T_L$, as a function of
the applied field obtained from our Monte Carlo code for a GaAs bulk
with $n=10^{16}$ cm$^{-3}$ at $T_L$=$300$ K. These results are
compared with the calculations from the KSBEs (see Fig. 15(b) of
Ref.\cite{Jiang09}). Our finding for the spin lifetime well agrees
with the theoretical results in all the investigated range, while
the values of the electron temperature are systematically slightly
lower than those obtained from the KSBE approach. Once more time,
this small deviation can be due to the fact that the Coulomb
scattering is neglected in our computation. In fact, the
electron-electron (e-e) scattering is very important in determining
the hot-electron temperature and the hot-electron temperature
influences both the electron-longitudinal optical (e-LO) phonon and
the electron-impurity scattering. Hence the e-e scattering
effectively influences the spin relaxation. However, since at high
temperatures, the e-e scattering is weak compared to e-LO phonon
scattering, it leads only to a marginal decrease in the spin
relaxation \cite{Zhang08}. Moreover, R\"{o}mer et
al.~\cite{Romer2010} have experimentally shown that, for doping
concentrations below the metal-to-insulator transition, the
electron-electron interaction is weaker at low impurity densities.
In our finding this decrease is surely negligible for the two lower
values of donor concentrations, namely $n=10^{13}$ cm$^{-3}$ and
$n=10^{15}$ cm$^{-3}$.

The analysis of the influence on the spin lifetime of the
intravalley and intervalley Coulomb scattering at high lattice
temperatures is a very interesting problem which deserves further
detailed investigations that will be the subject of a forthcoming
paper.

\section{Conclusions}\label{sect4}
A full understanding of the role played by the lattice temperature,
the doping density and the amplitude of high-intensity electric
field on the electron spin dynamics in semiconductors is essential
for the design and the fabrication of spintronic devices.

In this work we have studied the spin lifetimes of an ensemble of
conduction electrons, drifting in doped n-type GaAs bulk crystal, by
using a semiclassical Monte Carlo transport model. The spin
depolarization is investigated in a wide range of lattice
temperatures and doping densities by simulating the electrons driven
by an electric field with amplitude $0.1 < F < 6$ kV/cm, including
both the $\Gamma$-valley and the $L$-valleys into the spin dephasing
dynamics.

Our results show that the electron spin lifetime is not marginally
influenced by the driving electric field, the lattice temperature
and the impurity density, which hence represent key parameters into
the depolarization process. We find a significant reduction of the
average spin polarization lifetime at high values of the electric
field, caused by the stronger spin-orbit coupling of electrons in
the $L$-valleys. In the nondegenerate regime the doping density
considerably affects spin lifetimes at nitrogenum temperature and
weak intensity of the driving electric field. Moreover, for field
amplitudes greater than $2.5$ kV/cm, we observe spin lifetimes
longer at room lattice temperatures with respect to those observed
at $T_L=77$ K.

The prediction capability of our numerical code has been positively
validated through a comparison with the available experimental
results and with the calculations obtained by a different
theoretical approach.

\ack This work was partially supported by MIUR and CNISM-INFM. The
authors acknowledge CASPUR for the computing support via the
standard HPC grant 2010. Authors are very thankful to Prof. Dr. M.W.
Wu, who improved this work by precious comments and suggestions.


\section*{References}
\begin{thebibliography}{99}

\bibitem{FabianDasSarma1999}
Fabian J and Das Sarma S 1999 {\it J. Vac. Sci. Technol. B} {\bf
17}, 1708

\bibitem{Wolf2001}
Wolf S A, Awschalom D D, Buhrman R A, Daughton J M, Von Moln\'{a}r
S, Roukes M L, Chtchelkanova A Y and Treger D M 2001 {\it Science}
{\bf 294}, 1488

\bibitem{Awschalom2007}
Awschalom D D and Flatt\'{e} M E 2007 {\it Nature Phys.} {\bf 3},
153

\bibitem{Fabian2007}
Fabian J, Matos-Abiage A, Ertler C, Stano P and \v{Z}uti\'{c} I 2007
{\it Acta Phys. Slov.} {\bf 57}, 565

\bibitem{Lou07}
Lou X, Adelmann C, Crooker S A, Garlid E S, Zhang J, Madhukar Reddy
K S, Flexner S D, Palmstr{\o}m C J and Crowell P A 2007 {\it Nature
Phys.} {\bf 3}, 197

\bibitem{Jon07}
Jonker B T, Kioseoglou G, Hanbicki A T, Li C H and Thompson P E 2007
{\it Nature Phys.} {\bf 3}, 542

\bibitem{DyakonovEd}
Dyakonov M I (Ed.) 2008 {\it Spin Physics in Semiconductors}
(Berlin) Springer, Solid-State Sciences, Vol. 110, and references
therein.

\bibitem{Gul09}
Gulayev Y V, Zilberman P E, Panas A I and Epshtein E M 2009 {\it
Usp. Fiz. Nauk} {\bf 52} (4), 335

\bibitem{Flatte09}
Flatt\'{e} M E 2009 {\it Nature} {\bf 462}, 419

\bibitem{Dash2009}
Dash S P, Sharma S, Patel R S, De Jong M P and Jansen R 2009 {\it
Nature} {\bf 462}, 491

\bibitem{WuReport2010}
Wu M W, Jiang J H and Weng M Q 2010 {\it Physics Reports},
{\bf 493} 61.

\bibitem{Hagele98}
H\"{a}gele D, Oestreich M, R\"{u}hle W W, Nestle N and Eberl K 1998
{\it Appl. Phys. Lett.} {\bf 73}, 1580

\bibitem{Sanada2002}
Sanada H, Arata I, Ohno Y, Chen Z, Kayanuma K, Oka Y, Matsukura F
and Ohno H 2002 {\it Appl. Phys. Lett.} {\bf 81}, 2788

\bibitem{Sato04}
Sato Y, Takahashi Y, Kawamura Y and Kawaguchi H 2004 {\it Jpn. J.
Appl. Phys.} {\bf 43}, L230

\bibitem{Stephens04}
Stephens J, Berezovsky J, McGuire J P, Sham L J, Gossard A C and
Awschalom D D 2004 {\it Phys. Rev. Lett.} {\bf 93}, 097602

\bibitem{Crooker05}
Crooker S A and Smith D L 2005 {\it Phys. Rev. Lett.} {\bf 94},
236601

\bibitem{Furis2006}
Furis M, Smith D L, Crooker S A and Reno J L 2006 {\it Appl. Phys.
Lett.} {\bf 89}, 102102

\bibitem{Beck2006}
Beck M, Metzner C, Malzer S and D\"{o}hler G H 2006 {\it Europhys.
Lett.} {\bf 75}, 597

\bibitem{Ivchenko1990}
Ivchenko E L, Lyanda-Geller Y B and Pikus G E 1990 {\it Sov.
Phys.-JEPT}, {\bf 71}, 550

\bibitem{Wu2000}
Wu M W and Ning C Z 2000 {\it Phys. Status Solidi B} {\bf 222}, 523

\bibitem{Weng2003}
Weng M Q and Wu M W 2003 {\it J. Appl. Phys.} {\bf 93}, 410

\bibitem{Weng2004}
Weng M Q, Wu M W and Jiang L 2004 {\it Phys. Rew. B} {\bf 69},
245320

\bibitem{Zhang08}
Zhang P, Zhou J and Wu M W 2008 {\it Phys. Rev. B} {\bf 77}, 235323

\bibitem{Jiang09}
Jiang J H and Wu M W 2009 {\it Phys. Rev. B} {\bf 79}, 125206

\bibitem{Mishchenko2003}
Mishchenko E G and Halperin B I 2003 {\it Phys. Rev. B} {\bf 68},
045317

\bibitem{Saikin2004}
Saikin S 2004 {\it J. Phys.: Condens. Matter} {\bf 16}, 5071

\bibitem{Kiselev2000}
Kiselev A A and Kim K W 2000 {\it Phys. Rev. B} {\bf 61}, 13115

\bibitem{Bournel2000}
Bournel A, Dollfus P, Cassan E and Hesto P 2000 {\it Appl. Phys.
Lett.} {\bf 77}, 2346

\bibitem{Barry2003}
Barry E A, Kiselev A A, Kim K W 2003 {\it Appl. Phys. Lett.} {\bf
82}, 3686

\bibitem{Saikin2003}
Saikin S, Shen M, Cheng M C and Privman V 2003 {\it J. Appl. Phys.}
{\bf 94}, 1769

\bibitem{Pramanik2003}
Pramanik S, Bandyopadhyay S and Cahay M 2003 {\it Phys. Rev. B} {\bf
68}, 075313

\bibitem{Shen2004}
Shen M, Saikin S, Cheng M C and Privman V 2004 {\it Mathematics and
Computers in Simulation} {\bf 65}, 351

\bibitem{Pershin2005}
Pershin Y 2005 {\it Phys. Rev. B} {\bf 71}, 155317

\bibitem{Saikin2006}
Saikin S, Shen M and Cheng M C 2006 {\it J. Phys.: Condens. Matter}
{\bf 18}, 1535

\bibitem{Yu2002}
Yu Z G and Flatt\'{e} M E 2002 {\it Phys. Rev. B} {\bf 66}, 201202;
2002 {\it Phys. Rev. B} {\bf 66}, 235302

\bibitem{Martin2003}
Martin I 2003 {\it Phys. Rev. B} {\bf 67}, 014421

\bibitem{Hruska2006}
Hru\v{s}ka M, Kos \v{S}, Crooker S A, Saxena A and Smith D L 2006
{\it Phys. Rev. B} {\bf 73}, 075306

\bibitem{Fu2008}
Fu J Y, Weng M Q and Wu M W 2008 {\it Physica E} {\bf 40}, 2890

\bibitem{Elliott54}
Elliott R J 1954 {\it Phys. Rev.} {\bf 96}, 266

\bibitem{Yafet63}
Yafet Y 1963 {\it Solid State Physics}, edited by F. Seitz and D.
Turnbull (New York) Academic, Vol. 14, p. 2.

\bibitem{Perel1971}
Dyakonov M I and Perel V I 1971 {\it Sov. Phys. - Solid State} {\bf
13}, 3023

\bibitem{Dyakonov2006}
Dyakonov M I 2006 {\it Physica E} {\bf 35}, 246

\bibitem{BAP1976}
Bir G L, Aronov A G and Pikus G E 1976 {\it Sov. Phys. - JETP} {\bf
42}, 705

\bibitem{Abragam61}
Abragam A 1961 {\it The Principles of Nuclear Magnetism}, (Oxford)
Clarendon Press.

\bibitem{Paget77}
Paget D, Lampel G, Sapoval B and Safarov V I 1977 {\it Phys. Rev. B}
{\bf 15}, 5780

\bibitem{Pikus89}
Pikus G E and Titkov A N 1989 {\it Optical Orientation}, edited by
Meyer F, (Leningrad) Nauka.

\bibitem{Litvi2010}
Litvinenko K L, Leontiadou M A, Li J, Clowes S K, Emeny M T, Ashley
T, Pidgeon C R, Cohen L F, Murdin B N 2010 {\it Appl. Phys. Lett.}
{\bf 96}, 111107


\bibitem{Jacoboni}
Jacoboni C and Lugli P 1989 {\it The Monte Carlo Method for
Semiconductor Device Simulation}, edited by S. Selberherr, (Wien)
Springer.

\bibitem{Tomizawa}
Tomizawa K 1993 {\it Numerical simulation of submicron semiconductor
devices}, (London, Boston) Artech House.

\bibitem{Moglestue}
Moglestue C 1993 {\it Monte Carlo Simulation of Semiconductor
Devices}, (London) Chapman and Hall.

\bibitem{Persano2000}
Persano Adorno D, Zarcone M and Ferrante G 2000 {\it Laser Phys.}
{\bf 66}, 310

\bibitem{Romer2010}
R\"{o}mer M, Bernien H, M\"{u}ller G, Schuh D, H\"{u}bner J and
Oestreich M 2010 {\it Phys. Rev. B} {\bf 81}, 075216

\bibitem{Krich2007}
Krich J J and Halperin B I 2007 {\it Phys. Rev. Lett.} {\bf 98},
226802

\end{thebibliography}
\end{document}